\documentclass[11pt]{article}
\usepackage{amssymb}

\usepackage[top=2.75cm, bottom=2.75cm, left=2.75cm, right=2.75cm]{geometry}

\def\x{\times}

\def\o+{\oplus}

\def\beqa{\begin{eqnarray}}
\def\eeqa{\end{eqnarray}}

\begin{document}
\thispagestyle{empty}
\rightline{LMU-ASC 12/09}
\vspace{2truecm}
\centerline{\bf \large From Local to Global in F-Theory Model Building}
\vspace{0.5truecm}

\vspace{0.5truecm}
\centerline{Bj\"orn Andreas$^1$ and Gottfried Curio$^2$}

\vspace{.6truecm}
\centerline{$^1${\em Institut f\"ur Mathematik, Humboldt Universit\"at zu Berlin}}
\centerline{\em Rudower Chaussee 25, 12489 Berlin, Germany}
\centerline{$^2${\em Arnold-Sommerfeld-Center for Theoretical Physics}}
\centerline{{\em Department f\"ur Physik, Ludwig-Maximilians-Universit\"at M\"unchen}}
\centerline{{\em Theresienstr. 37, 80333 M\"unchen, Germany}}

\vspace{1.0truecm}

\begin{abstract}

When locally engineering $F$-theory models some $D7$-branes
for the gauge group factors are specified and matter is localized on the intersection curves of the compact parts of the world-volumes. In this note we discuss to what extent one can draw conclusions about $F$-theory models by just restricting the attention locally to a particular seven-brane.
Globally the possible $D7$-branes are not independent from each other and the (compact part of the)
$D7$-brane can have unavoidable intrinsic singularities. Many special intersecting loci which were not chosen by hand occur inevitably,
notably codimension three loci which are {\em not} intersections of matter
curves. We describe these complications specifically in a global $SU(5)$ model and also their impact on the tadpole cancellation condition.

\end{abstract}

\newpage

\section{Inevitable Effects in Global F-Theory}

We consider F-theory on ${\bf R^{3,1}}\times X$
where $X$ is an elliptically fibered Calabi-Yau fourfold over a complex threefold $B_3$. This set-up is used to describe in an effective fourdimensional theory certain gauge theories with matter.\\

\noindent{\em The Local Point of View}

\noindent
Engineering a GUT model in local F-theory models
essentially the following procedure is chosen
\begin{itemize}
\item {\em Codimension one:}
One demands the existence of a $D7$-brane (the compact part of its world-volume is a divisor $D_G$ in $B_3$) which encodes the GUT group $G$. This is assumed to give just the gauge group.
\item
{\em Codimension two:}
Then one demands the existence of further $D7$-branes (this again refers to codimension one) such that at the intersection curves in $D_G$ (we are speaking here of the compact parts of the
world-volumes) those enhancements occur which give the matter multiplets (quarks, leptons, Higgses)
one wants to encode.
\item
{\em Codimension three:}
The matter curves in $D_G$ intersect at points.
One tries to arrange things such that at certain points intersections occur which encode suitable Yukawa couplings.
\end{itemize}

So these three steps refer, geometrically speaking, to more and more
specialized choices in (complex) codimension-one, two and three in $B_3$.
The underlying philosophy is that by restricting attention to what happens
inside $D_G$, one stays decoupled from 'global complications'.
Technically the decoupling of the surrounding $B_3$ (and of gravity) is achieved
by taking the size of $B_3$ to infinity while holding the size of $D_G$ fixed,
or, reversing it, by making sure that $D_G$ is shrinkable in a fixed $B_3$
(this leads to $D_G$ being a del Pezzo surface) [\ref{V1}].
This suggests the idea that the effective theory has decoupled itself thereby from
the 'complications of global geometry' and can that one can focus on local considerations.
This philosophy came from the $D$-brane models where string theory is used
to actually doing field theory. The reason for the whole procedure is that the engineered field theory models are 'better' than ordinary field theory models; this is because at the end one nevertheless still wants to participate at the benefits of having a field theory which is thought to be embeddable in a full string theory.
So the justification of the whole enterprise stands and falls with this
embeddability in a consistent global string model, i.e.,~the hope is a possible passage from local to global.\\

\noindent
{\em The Global Point of View}

There were however always signs that global consistency has to be taken into
account even in a local procedure. The $D3$-brane charge tadpole
cancellation condition showed already that one has to have always an eye
on {\em all} global contributions, cf.~section~\ref{tadpole section} below.

One feature which makes F-theory models particularly interesting is that
a similar story occurs also for the $D7$-branes:
the condition that the $D7$-branes with compact parts of world-volumes
$D_i$ (and multiplicities $m_i$) satisfy
\beqa
\label{cohom discr equ}
\sum m_i D_i &=& 12 c_1(B_3).
\eeqa
This is the cohomological decomposition of the discriminant $\{\Delta=0\}$.

So the following has to be taken into account. First there is the global binding of $D7$-branes equ. (\ref{cohom discr equ}).
Furthermore, when engineering the matter curves that one wants to have
from certain surface components, all topological unavoidable intersections with $D_G$ must be taken into account.
Especially important in this respect is the divisor $D_{I_1}$ of an $I_1$-fiber singularity. This is the locus where just the fiber degenerates simply and thus makes
itself not felt as a singularity of the total space\footnote{In the 8-dimensional case of $F$-theory on $K3$
the collision of $n$ nodal fibers in the base ${\bf P}^1$
leads to an $I_n=A_{n-1}$ singularity of the total space which results in
a non-abelian gauge group in an effective theory.}. As we will recall below in our main example of $G=SU(5)$ it is the intersection
of $D_G$ with this $D7$-brane which leads to the matter we want to engineer.

\begin{itemize}
\item
{\em Codimension-One:} Starting with the $D7$-brane wrapping $D_G$ gives effectively the gauge group $G$.
Because of the global consistency relation equ. (\ref{cohom discr equ})
it is usually {\em not} possible to encode just a {\it pure} gauge theory!
This means that even if one does not want to encode further non-abelian
gauge group factors nevertheless one has to satisfy equ. (\ref{cohom discr equ}), at least via the contribution of the $I_1$ surface $D_{I_1}$.
The latter will usually for topological reasons inevitably intersect $D_G$
and give matter curves. Let us consider for illustration the case where $B_3$ is itself ${\bf P^1}$ fibered over a surface $B_2$ (such that $X$ is
$K3$ fibered, the case with a heterotic dual; the fibration type of the ${\bf P^1}$ over $B_2$ is encoded by a cohomology class $t$ in the base\footnote{cf.~the classification by a number $n$ when having a ${\bf P^1}$ instead of $B_2$ and a Hirzebruch surface instead of $B_3$}). If $G$ is in the $E$ series only {\em one} matter curve occurs, whose
cohomology can be 'turned off' by choosing $t$ related in a specific way to
$c_1(B_2)$; this means that the two surfaces can be geometrically separated.
In the $A$ and $D$ series, however, {\em two} matter curves occur with different cohomology class which can not be 'turned off' both at the same
time; that is the two surfaces can not be separated from each other, we get
{\em inevitably} one matter curve at least just from consistency.

Furthermore, this unavoidable $D_{I_1}$ component turns out to be a {\em singular} surface, again something which was not chosen but
which rather just occurs in the detailed consideration.
It has always a curve of intrinsic cusp singularities, for example,
which is (accidentally) at the same time
a curve of cuspidal fiber singularities, i.e.,~the locus in
$D_{I_1}$ where the fiber type changes from nodal $I_1$ to cuspidal $II$.

In certain cases, like $SU(n)$ for $n=4$ or $5,6$,
there is a further curve of intrinsic singularities built by
tacnodes or higher double points; 'accidentally' it happens that this curve lies even in $D_G$, {\it so it is actually one of the matter curves}
which thus occurs here in $D_G$ in a collision with a singular locus of the second $D7$-brane.
\item
{\em Codimension two:}
Therefore we have the following situation in codimension two: although one might choose specializations which give other components of the discriminant and lead to matter curves, we will
usually have the remaining amount of the discriminant divisor $(\Delta)$
which represents the $I_1$-surface $D_{I_1}$. Except for a very special
``{\em separation case}'' (where $D_{I_1}$ and $D_G$ are disjoint)
this leads to matter curves at the intersection loci in $D_G$.
We may have chosen at will certain $D7$-branes already; our point here is that because of the 'global binding' equ. (\ref{cohom discr equ}) we usually get further $D7$-branes, i.e.~components
$D_i$ of $(\Delta)$, which we possibly did not want to have and which nevertheless
lead to further intersection curves in $D_G$, i.e.~further matter
multiplets in the effective four-dimensional effective theory. That is, even by restricting attention to $D_G$ and to some intersection curves
we want to engineer, we are not protected against further intersections
from further $\Delta$-components; rather we have the overall global binding equ.~(\ref{cohom discr equ}) to satisfy. The minimalistic way is then the further $D_{I_1}$ component (which luckily often gives just the relevant matter already).

If one wants a matter curve not contained in $D_G$ this must arise
from further surface components $D_i$ of $(\Delta)$ which in turn will
interesect $D_G$.

$D_{I_1}$ has singular curves (even in $D_G$) which signify further special behaviour, cf.~above.

For phenomenological reasons (as we will see in the $SU(5)$ example)
we may want to have besides the fermionic matter of quarks and leptons
also Higgses; these come from the same type (fundamental)multiplet type as one of the matter curves; so the corresponding curve has to be reducible, a condition in complex structure moduli which has to be stabilized.

\item {\em Codimension three:}
Finally there are points of special further enhancements.
Among them are the interesections of two matter curves in $D_G$
(whereas a meeting of three curves usually has to be stabilised), but {\em there
are more special enhancement points as the global analysis reveals}
(notably the $P=Q=0$ locus in the $SU(n)$ cases). {This is an important global feature}.
Furthermore the intrinsic cusp curve $C$ of
$D_{I_1}$ intersects $D_G$ in some points\footnote{over $C$ lie
also more complicated (cuspidal) fibers
which collide at these points with the $G$-singularity; also
various complicated point singularities
of $D_{I_1}$, detected by an analysis of the discriminant equation,
can occur}. \end{itemize}

The message of these details is simple. If one
wants to build a GUT model with a specific matter content and wants to draw specific conclusions, one has reason to care about the global structure of the discriminant (the $D7$-brane components)
and the finer specialization structures of this locus. Surely when it comes to tadpole cancellation
a detailed overview of all possible contributions
to the Euler number is required. If we assume that actually the resolution of the singular model
is concerned here (and if that exists), one has to do
all the needed blow-up processes, cf.~the cases in [\ref{AC}].
These matters are not yet fully elucidated. One case is discussed
in [\ref{Aluffi}] and we will give also a discussion below.

{\em Note added:} As this note was prepared for final publication
the paper [\ref{HKTW}] appeared in which also codimension three loci
in the $SU(5)$ model were investigated. Related papers are [\ref{DW}].

\section{The Discriminant Equation and Singularity Loci}

In this section we recall some details of the geometry of the $F$-theory models. We assume the existence of a section $\sigma\colon B_3\to X$ and that $X$ can be described by a Weierstrass model
\beqa
\label{Weier}
y^2&=&x^3+fx+g
\eeqa
where $f$ and $g$ are sections of $K_{B_3}^{-4}$ and $K_{B_3}^{-6}$, respectively. The elliptic fiber degenerates over the discriminant locus $D=\{\Delta=0\}$ of the above equation, where
\beqa
\label{discrim equat}
\Delta=4f^3+27g^2.
\eeqa

We will denote the cohomology classes of the vanishing divisors $(f)$ and $(g)$ by $F:= 4 c_1(B_3)$ and $G:= 6 c_1(B_3)$; similarly $D:=12 c_1(B_3)$ for $(\Delta)$.
For $p\in D\subset B_3$ the type of singular fiber is determined
by the orders of vanishing $a:=ord(f),
b:=ord(g)$ and $c:=ord(\Delta)$
according to the Kodaira list of singularities of elliptic fibrations.

{\footnotesize \begin{center}
\begin{tabular}{|c|c|c|c|c|}
\hline
$a$ & $b$   & $c$ & $\rm{fiber}$  & \rm{singularity} \\ \hline
$\ge 0$ & $\ge 0$          & $0$    & $smooth$ & $none$      \\    \hline
$\ \ \  0$ & $ \ \ \ 0$      & $n$        & ${I}_n$  & ${A}_{n-1}$\\\hline
$\ge 1$ & $ \ \ \ 1$        & $2$      & ${II}$   & $none$    \\  \hline
$\ge 1$ & $\ge 2$          & $3$    & ${III}$  & $A_1$     \\ \hline
$\ge 2$ & $\ \ \ 2$        & $4$    & ${IV}$     & $A_2$      \\  \hline
$\ \ \ 2$ & $\ge 3$        &$n+6$     & ${ I}_n^*$  & $D_{n+4}$ \\ \hline
$\ge 2$ & $\ \ \ 3$        &$n+6$   & ${ I}_n^*$  & $D_{n+4}$ \\  \hline
$\ge 3$ & $\ \ \ 4$        & $8$    & ${IV}^*$   & $E_6$       \\ \hline
$ \ \ \ 3$ & $\ge 5$        & $9$      & ${ III}^*$  & $E_7$      \\  \hline
$\ge 4$ & $\ \ \ 5$        & $10$   & ${ II}^*$   & $E_8$     \\
\hline
\end{tabular}
\end{center}}
So for the singular fiber type $I_1$ (nodal) or
type $II$ (cuspidal) no singularity of the total space arises.
This list originated in the case corresponding to an $F$-theory model
on an elliptically fibered $K3$ surface, i.e.,~a compactification to 8
dimensions; indicated are the vanishing orders in the coordinate
$z$ in the base ${\bf P^1}$ of the fibration.
This type of structure will be prolonged adiabatically in the following
to compactification models of (6 or) 4 dimensions. The complex threedimensional base $B_3$ of the elliptically fibered
Calabi-Yau fourfold $X$ will be assumed to be ${\bf P^1}$ fibered
over an own base surface $B_2$
(equivalently $X$ is assumed to be fibered by elliptic $K3$ surfaces
over $B_2$; this case has a heterotic dual).

So we consider $B_3$ being
a ${\bf P^1}$ bundle which is the projectivization ${\bf P}(Y)$ of a vector bundle $Y={\cal O}\oplus {\cal T}$ with ${\cal T}$ a line bundle over $B_2$ and ${\cal O}={\cal O}_{B_2}$. Furthermore, let ${\cal O}(1)$ be a line bundle on the total space of ${\bf P}(Y)\rightarrow B_2$ which restricts on each ${\bf P^1}$ fiber to the corresponding line bundle over ${\bf P^1}$. With
$r=c_1({\cal O}(1))$, $t=c_1({\cal T})$ and $c_1({\cal O}\otimes {\cal T})
=r+t$ then the cohomology ring of $B_3$ is generated over the cohomolgy ring of $B_2$ by the element $r$ with the relation $r(r+t)=0$. The total Chern class $c(B_3)=c(B_2)(1+r)(1+r+t)$ gives (we set $c_1:=c_1(B_2)$; here $c_1$ and $t$
are understood as pullbacks to $B_3$)
\beqa c_1(B_3)&=& c_1+2r+t, \ \ \ \ c_2(B_3)=c_2+c_1t+2c_1r.
\eeqa

\noindent
{\em Codimension-One Loci}

If we engineer an $ADE$ gauge group $G$ in four dimensions we just demand a corresponding surface component in $D$. Let us call this surface component $B_2$ and denote its cohomology class by $r$.
The next step in the engineering process is to demand some loci where
matter charged under $G$ is located; so let us look what happens inevitably in a global model when one starts and demands just the component $B_2$. We will have automatically the decomposition $D=D_1+D_2$
where $D_1$ denotes the component with generic $I_1$ fibers and $D_2$ has $G$ fibers. This leads to the cohomological relations
$F_2=ar,
G_2=br,
D_2=cr$
(for $f=f_1f_2$ and $g=g_1g_2$ with $f_2=z^a, g_2=z^b$
where $z$ is the coordinate on the fiber ${\bf P^1}$ of $B_3$).
For the remaining locus $D_1$ of $I_1$ fibers and the other terms one gets
\beqa
\label{divisor1 equations}
F_1&=&4c_1+(8-a)r+4t\nonumber\\
G_1&=&6c_1+(12-b)r+6t\\
D_1&=&12c_1+(24-c)r+12t.\nonumber
\eeqa
For $E_k$-singularities
the $I_1$-surface component $D_1$ is given
by the equation $4f_1^3-27g_1^2=0$.
For $SU(n)$, however, we have $f=f_1, g=g_1$ and one can split off in $\Delta$ a $z^n$-factor
\beqa
\label{discriminant terms}
\{4f_1^3+27g_1^2=0\}&=&D_1+nr
\eeqa
For $SO(2(n+4))$ we define $f_2=z^2, g_2=z^3$ and get
again equ.~(\ref{discriminant terms}).

\noindent
{\em Codimension-Two Loci}

The intersection curves between the $G$-surface $D_2$
and the inevitably occurring $I_1$-surface $D_1$ have an
interpretation as matter locations
because the collision of singularity types will lead
to a gauge group enhancement: for example, if $G=SU(n)$ is enhanced
to $SU(n+1)$ we get a fundamental representation from the decomposition
${\bf ad}_{SU(n+1)}={\bf V}\oplus \overline{{\bf V}}\oplus {\bf ad}_{SU(n)}
\oplus {\bf C}$
under $SU(n)\x U(1)\subset SU(n+1)$,
or an antisymmetric one from an $SO(2n)$ enhancement
${\bf ad}_{SO(2n)}={\bf \Lambda^2 V}\oplus {\bf \Lambda^2 \overline{V}}
\oplus {\bf ad}_{SU(n)}\oplus {\bf C}$.
Similarly we get the ${\bf V}$ and the ${\bf S}$ of $SO(10)$ from
the decompositions of enhancements
${\bf ad}_{SO(12)}={\bf V}\oplus \overline{{\bf V}}\oplus {\bf ad}_{SO(10)}
\oplus {\bf C}$ and
${\bf ad}_{E_6}={\bf S}\oplus {\bf \overline{S}}\oplus {\bf ad}_{SO(10)}
\oplus {\bf C}$.
Likewise an $E_7$ enhancement of $E_6$ will provide the ${\bf 27}$.

Let us look at the first few non-trivial cases
(we give in the last three entries the dual heterotic data
where an $H_{V}\x E_8$ bundle $(V,V_2)$ is given with $H_{V}=SU(n)$,
cf.~section~\ref{SU(5) section})
\begin{center}
\begin{tabular}{|c|ccc|c|cc|ccc|}
\hline
$G$    &$a$&$b$& $c$ & matter curve(s) & $fib_{enh}$&
 matter&$H_{V_1}$&het&{\rm het. loc.}\\
\hline
$E_7$  &$3$&$5$& $9$ &$f_{4c_1-t}$ &"$E_8$" &
$(\frac{1}{2}){\bf 56}$&$SU(2)$&$H^1(Z,V)$&$a_2$\\ \hline
$E_6$  &$3$&$4$& $8$ &$q_{3c_1-t}$&$E_7$&${\bf 27}$&$SU(3)$&$H^1(Z,V)$&
$a_3$\\ \hline
$SO(10)$&$2$&$3$& $7$ &$h_{2c_1-t}$&$E_6$&${\bf 16}$&$SU(4)$&$H^1(Z,V)$&$a_4$\\
$$     &$$ &$$ & $$&$q_{3c_1-t}$&$SO(12)$&${\bf 10}$&
$$&$H^1(Z,\Lambda^2V)$&$a_3$\\ \hline
$SU(5)$  &$0$&$0$& $5$ &$h_{c_1-t}$ &$SO(10)$&${\bf 10}$&$SU(5)$&
$H^1(Z,V)$&$a_5$\\
$$&$$&$$&$$&$P_{8c_1-3t}$&$SU(6)$&${\bf \bar{5}}$ &$$
&$H^1(Z,\Lambda^2V)$&$R(a_i)$\\
\hline
\end{tabular}
\end{center}

Some further matter curves are given here in the following table
where the polynomial, including multiplicities,
giving the defining equation of $D_1r$ is displayed
\begin{center}
\begin{tabular}{|c|c|}
\hline
$G$   & equ.\, of \, $D_1r$  \\
\hline
\hline
$A_1$ & $H_{2c_1-2t}^2P_{8c_1-6t}$ \\
\hline
$A_n$ & $h_{c_1-t}^4P_{8c_1-(7-n)t}$ \\
\hline
$D_4$ & $\prod_{i=0}^2 (h_{2c_1-t}^2+\omega^i P_{2c_1-t}^2)$ \\
\hline
$D_5$ & $h_{2c_1-t}^3 q_{3c_1-t}^2$ \\
\hline
$D_6$ & $h_{2c_1-t}^2 P_{4c_1-t}^2$ \\
\hline
$E_k$ & $q_{\frac{12}{k'}c_1-t}^{k'}$\\
\hline
\end{tabular}
\end{center}
where $n=2,3,4,5$, $\omega=e^{2\pi i/3}$, $k=6,7,8$ and $k'=10-k$.
In all cases we get the correct sum for the total cohomology class
\beqa
\label{total curve class}
D_1 r&=& \Big(12c_1-(12-c)t\Big)r.
\eeqa

\noindent{\em Separation Cases and Pseudo-Separation Cases}

If we want to obtain a degeneration which is purely in codimension one, we must arrange things such that $D_1$ and $r$ do not intersect. This can be achieved by adjusting the Chern class $t$ which specifies how the ${\bf P^1}$ is fibered over $B_2$. The table shows that for $D_n$ and $A_n$ there is more than one matter curve, so we can not 'turn off'
cohomologically all of them simultaneously. However, for the $E$ series
this is possible as there only one matter curve appears.
From (\ref{total curve class}) the separation of $D_1$ 
and $r$ can be achieved for $t$ given by
\beqa
t={12\over{12-c}}c_1,
\eeqa
provided the right hand side is an integral class. Therefore for the $E_n$ series with $c=8,9,10$ we can adjust a 'separation case' between $D_1$ and $r$, i.e.~a matter free situation, by setting $t=3c_1, 4c_1, 6c_1$ for $E_6, E_7, E_8$ (the case $E_8$ is somewhat special, cf.~[\ref{KLRY}]).
In the $D$ series we find only for $D_4$ a realizable codimension-one case, with $t=2c_1$. In the $A$ series only a pseudo-separation case can be established by setting $t=c_1$, i.e.~the matter can not be completely 'turned off', only one of the two matter curves is turned off cohomologically,
say $(h)$.\\

\noindent{\em The Cusp Curve and Further Singular Curves}

Inevitably occurs another relevant codimension two locus:
the cusp curve. The {\em naive} cusp locus is $C_{naive}=\{f_1=0=g_1\}$.
In case (\ref{discriminant terms}) applies this naive locus will contain also higher singularities over the matter curve $h$ such that
the true cusp set is \beqa
C&=&C_{naive}-x(h),
\eeqa
where $x$ is the intersection multiplicity of $f_1$ and $g_1$ along $(h)$
(computed via their resultant).
The cuspidality means here that not only $C$ is a locus of intrinsic cusp singularities of $D_1$
but the singularity type of the elliptic fiber
over points in $C$ is also cuspidal ($y^2-x^3=0$).

We will have further curves of intrinsic singularites of $D_1$ [\ref{AC}]: for $SU(4)$ a
curve of tacnodes, for $SU(5)$ and $SU(6)$ a curve of higher double points. There one needs two or more blow ups.\\

\noindent{\em Codimension-Three Loci}

For $G=SU(n)$ or $SO(2n)$ one gets two matter curves
which intersect in some points of $B_2$ (considered in
detail below for $SU(5)$).
Two other possible types of codimension three loci are
point singularities of $D_1$ and intersection points of the cusp curve $C$ with $B_2$.
Further there occurs a codimension three locus $(P)\cap (Q)$, cf.~below,
for global reasons.

\section{\label{SU(5) section}An $SU(5)$ GUT Model}

We start from the Weierstrass model (\ref{Weier}) and expand $f$ and $g$ in the section given by $z$, the coordinate of the ${\bf P^1}$ fiber
of $B_3$ over $B_2$; so $z=0$ corresponds to the locus $B_2$ of $SU(5)$ GUT group (the divisor $r$). Note that the cohomology class
$4c_1(B_3)$ of the bundle of which $f$ is a section
reads on $r$ just $4(c_1-t)$. Now develop $f$
in a polynomial in $z$ with coefficient functions given by suitable
sections over $B_2$. The constant term has precisely the mentioned cohomological 'degree' $4(c_1-t)$;
each $z$-power then consumes one $-t$ from this class because
the vanishing divisor of the section $z$ is again $r$ and we have
$r|_r=-t|_r$; therefore the coefficient of $z^i$ is some $f_{4c_1-4t+it}$
\beqa
f&=&\frac{1}{2^4\cdot 3}\sum_{i=0}^7 f_{4c_1-(4-i)t}\, z^i + {\cal O}(z^8),\\
g&=&\frac{1}{2^5\cdot 3^3}\sum_{j=0}^7 g_{6c_1-(6-j)t}\, z^j+{\cal O}(z^8),
\eeqa
where the $f_{4c_1-(4-i)t}$, $g_{6c_1-(6-j)t}$ are sections of line bundles over $B_2$ with Chern classes indicated by the subscripts.
As we will be interested only in the development of the discriminant
$\Delta$ up to the order $z^7$ we keep only the terms shown. Actually we will take as highest terms for $f$ and $g$ the terms $f_{4c_1}z^4$ and $g_{6c_1}z^6$, respectively (note that these terms
already correspond heterotically to the complex structure moduli
of the heterotic elliptic Calabi-Yau threefold, whereas the lower terms
will be relevant for description of the first $E_8$ bundle).
So the actual starting point will be
\beqa
f&=&\frac{1}{2^4\cdot 3}
\Big(f_{4c_1-4t}+f_{4c_1-3t}\, z+f_{4c_1-2t}\, z^2+f_{4c_1-t}\, z^3
+f_{4c_1}\, z^4\Big),\\
g&=&\frac{1}{2^5\cdot 3^3}\Big(g_{6c_1-6t}+g_{6c_1-5t}\, z+g_{6c_1-4t}\, z^2
+g_{6c_1-3t}\, z^3+g_{6c_1-2t}\, z^4+g_{6c_1-t}\, z^5+g_{6c_1}\, z^6\Big).
\;\;\;\;\;\;
\eeqa
The discriminant expression $\Delta$ in equ.~(\ref{discrim equat})
will now also be expanded as a polynomial in $z$ where for an $I_5=A_4$ singularity the coefficients of $z^i$ for $i=0,1,2,3,4$
have to cancel, giving expressions for $f_{4c_1-(4-i)t}$ and $g_{6c_1-(6-j)t}$
(subscripts indicate the 'cohomological degrees')\footnote{in general one has to invoke a generalised Weierstrass/Tate equation [\ref{BIKMSV}];
the dictionary to the coefficients of [\ref{HKTW}], for example, is
$a_5=h, -4a_4=H, 12 a_3=q, 48 a_2=f_{4c_1-t}, 48 f_0=f_{4c_1},
-288a_4f_0+864a_0=g_{6c_1-t}, 864g_0=g_{6c_1}$}
\beqa
f_{4c_1-4t}&=&-h^4\\
f_{4c_1-3t}&=&2h^2H\\
f_{4c_1-2t}&=&2hq-H^2\\
g_{6c_1-6t}&=&h^6\\
g_{6c_1-5t}&=&-3h^4H\\
g_{6c_1-4t}&=&3h^2(H^2-hq)\\
g_{6c_1-3t}&=&\frac{3}{2}h(2Hq-hf_{4c_1-t})-H^3\\
g_{6c_1-2t}&=&\frac{3}{2}(f_{4c_1-t}H+q^2 -h^2f_{4c_1})
\eeqa
Here we introduced arbitrary sections of the following cohomological degrees
\beqa
h_{c_1-t}, \;\;\; H_{2c_1-t}, \;\;\; q_{3c_1-t}, \;\;\; f_{4c_1-t}, \;\;\; g_{6c_1-t}
\eeqa
and similarly also $f_{4c_1}$ and $g_{6c_1}$.
The discriminant has then the following structure
\beqa
\label{SU(5) discriminant}
\Delta=c \, z^5 \Delta_1&=&c \, z^5\Big( h^4 P+h^2\Big[-2HP+hQ\Big]z
+\Big[-3q^2H^3+{\cal O}(h)\Big]z^2
+{\cal O}(z^3)\Big)
\eeqa
where $c= (2^{10}\cdot 3^3)^{-1}$ and \beqa
\label{P def}
P=P_{8c_1-3t}&=&-3Hq^2 -3f_{4c_1-t}qh+\Big[2g_{6c_1-t}-3f_{4c_1}H\Big]h^2\\
Q=Q_{9c_1-3t}&=&-q^3
-\Big(\frac{3}{4}f_{4c_1-t}^2+\Big[2g_{6c_1-t}-3f_{4c_1}H\Big]H\Big)h
+\Big(2g_{6c_1}h-3f_{4c_1}q\Big)h^2.
\eeqa
Let us now read off the various relevant subloci from the discriminant equation (\ref{SU(5) discriminant}).
\begin{itemize}
\item {\em Codimension one:} We encoded the divisor of $SU(5)$ singularity type by the factor $z^5$;
this is just the surface $B_2$ (of class $r$ and with multiplicity $5$).
The remaining factor (in curly brackets) gives the defining polynomial of $D_{I_1}$.

\item {\em Codimension two:} There are two matter curves, corresponding to singularity enhancements in codimension two in $B_3$, given by the loci $h=0$ ($SO(10)$ enhancement), leading to antisymmetric matter in the ${\bf 10}$ and ${\bf \overline{10}}$, and $P=0$ ($SU(6)$ enhancement), leading to fundamental matter in the ${\bf 5}$ and ${\bf \overline{5}}$ \beqa h=0 &\Longrightarrow &
(a,b,c)=(2,3,7) \;\;\buildrel \wedge \over = \;\;
A_4\to D_5 \\
P=0 &\Longrightarrow &(a,b,c)=(0,0,6) \;\; \buildrel \wedge \over = \;\;
A_4\to A_5.
\eeqa

As the Higgs fields $H_u$ and $H_d$ of the MSSM sit in the ${\bf 5}$
and ${\bf \overline{5}}$, respectively, we may want to have further
independent curves giving an $SU(6)$ enhancement like $P$. So we have
to tune the complex structure moduli of $P$ in such a way that the locus $P=0$ in $r$ becomes reducible and decomposes actually
in three curves (this may or may not come from a reducibility of the $I_1$-surface $D_1$ itself). Whether such a locus in the complex structure moduli space is somehow stabilized remains an open question.

\item {\em Codimension three:} The singularity type is enhanced even further in codimension three. Various such point loci occur in the $SU(5)$-surface $r$
(by equ.~(\ref{P def}) the intersection locus $h=P=0$ of the matter curves
is contained in either $h=H=0$ or $h=q=0$)
\beqa
h=H=0 &\Longrightarrow &(a,b,c)=(3,4,8) \;\;\buildrel \wedge \over = \;\; A_4
\to E_6 \\
\label{D6 points}
h=q=0 &\Longrightarrow &(a,b,c)=(2,3,8) \;\;\buildrel \wedge \over = \;\;
A_4\to D_6 \\
\label{A6 points}
P=Q=0 \;\;\;(\mbox{but not} \; h=q=0) &\Longrightarrow &(a,b,c)=(0,0,7) \;\;
\buildrel \wedge \over = \;\;A_4\to A_6
\eeqa
In equ.~(\ref{A6 points}) the conditions $P=Q=0$ have to be taken in a generic sense (the conditions $h=q=0$ in equ.~(\ref{D6 points}) imply also $P=Q=0$, but this is a non-generic solution).
Note that while the $E_6$ and $D_6$ enhancement points are expected
from a local ansatz as intersection of the matter curves $h$ and $P$,
the $A_6$ enhancement locus $(P)\cap (Q) - (h)\cap (q)$ arises from the precise structure of the discriminant equ.~(\ref{SU(5) discriminant}) in the global set-up.
\end{itemize}

\noindent
{\em Intrinsic Singularities of the $I_1$-Surface: The Cusp Curve $C$}

The surface $D_1$ has some intrinsic singularities. First equ.~(\ref{discrim equat}) suggests that the curve $\{f=g=0\}$ in $\{\Delta=0\}$ is a curve of intrinsic cusp singularities of $D_1$
(resolvable by one blow-up).
Actually by equ.~(\ref{Weier}) over the corresponding points
also lie cuspidal fibers (type $II$).

Actually the curve $\{f=g=0\}$ is reducible as $\{f=0\}$ and $\{g=0\}$
have the curve $\{h=0\}$ in $B_2$ as common component; on the level of divisors we have $(f)|_r=4(h)$ and $(g)|_r=6(h)$ from
\beqa
\label{f expansion}
f&=&-h^4+2h^2Hz+(2hq-H^2)z^2+f_{4c_1-t}z^3+f_{4c_1}z^4\\
g&=&h^6-3h^4Hz+3h^2(H^2-hq)z^2+\Big(\frac{3}{2}h(2Hq-hf_{4c_1-t})
-H^3\Big)z^3\nonumber\\
\label{g expansion}
&&+\frac{3}{2}\Big(f_{4c_1-t}H+q^2-h^2f_{4c_1}\Big)z^4
+g_{6c_1-t}z^5+g_{6c_1}z^6
\eeqa

The intersection multiplicity of $(f)=\{f=0\}$ and $(g)=\{g=0\}$ at the curve $(h)=\{h=0\}$ can be computed as the $h$-order of the resultant of the polynomials in $z$ given by $f$ and $g$; this gives the order $15$
(where $15=3n=ord_h Res(f,g)$ for the case of $I_n$ with $n\in \{4,5,6\}$).
Thus from the naive locus $(f)(g)$ a component $15(h)r$ has to be split off to get the true cusp curve
\beqa
\label{Cusp}
C&=&(f)(g)-15(h)r
\eeqa
(on the divisorial level in $D_1$). We find that cohomologically \beqa
\label{Cr points}
Cr&=&24[h]^2+15[h]t=3[h]\Big(8[h]+5t\Big)=3[h][P]=3[h]\Big([H]+2[q]\Big)
\eeqa
where the factor $3={\rm gcd}(24, 15)$ is a divisorial multiplicity in $C|_r$
such that $\#(C\cap r)=Cr/3$ as cardinality. Whether actually also
$C\cap r \subset (h)\cap (P)$ as point sets is less clear at first as we did in equ.~(\ref{Cr points}) just a cohomological computation. That $C\cap r\subset (h)$, however, is clear as the
cardinality is turned off by setting the cohomology class $[h]$ of $(h)$ in (\ref{Cr points}) to zero (or because actually {\em the divisor} $(h)$ occurs as factor in $C|_r$). We will show now that even $C\cap r\subset (h)\cap (P)=\Big((h)\cap (H)\Big)\cup \Big((h)\cap (q)\Big)$; possibly one has even $C\cap r= (h)\cap (P)$.
\vskip 0.3cm
\noindent
{\em Investigation of the Locus $C\cap B_2$}

Concerning the point locus $C\cap r$ think of $H$ (or $q$) and $h$ as local functions in the $r$-plane around a point $p\in C\cap r$; we know that $h(p)=0$ and we want to show that either $H(p)=0$ or $q(p)=0$.
If $x$ and $y$ are local coordinates near $p$ in $B_2$ then $C$ is parametrized as $(x(\tau), y(\tau), z(\tau))$ for some parameter $\tau$; we will take $h, H$ and $q$ locally as functions of $(x, y)$ and make the normalisation
that $\tau=0$ gives $z=0$ (this means that $z_0=0$ below). We saw above that the points in $Cr$ come with multiplicity $3$
and not $1$ so we cannot take $z$ itself as a parameter. Therefore we make the following ansatz for the parametrization of $C$ near $p$
(we have $h_0=0$, i.e.~$j>0$, from $p\in (h)$; similarly $p\in (H)$ just if $H_0=0$, and correspondingly for $(q)$)
\beqa
\label{Ansatz h}
h=h(\tau)&=&h_j\tau^j+h_{j+1}\tau^{j+1}+\dots\\
\label{Ansatz H}
H=H(\tau)&=&H_0+H_1\tau+H_2\tau^2+\dots\\
\label{Ansatz z}
z=z(\tau)&=&z_k\tau^k + z_{k+1}\tau^{k+1}+\dots
\eeqa
Here the ansatz for $z(\tau)$ comes with $k\geq 3$ from the multiplicity $3$. With this ansatz $f, g$ and $\Delta_1$ (the defining polynomial of $D_{I_1}$, i.e.~the factor in curly brackets in equ.~(\ref{SU(5) discriminant}))
become expressions in $\tau$ which by $C\subset (f)\cap (g)$ and $C\subset (\Delta_1)$ must vanish identically in $\tau$. As the coefficients of the individual $\tau$-powers must vanish let us look for the lowest order term. Note first that \beqa
\Delta_1&=&\Big[-3Hq^2+{\cal O}(h)\Big]
\Big(h^4-2h^2Hz+H^2z^2\Big) + {\cal O}(z^3)
\eeqa
from $P=-3Hq^2+{\cal O}(h)$.
We want to show that at an intersection point in $C\cap r$ either
$H_0=0$ or $q_0=0$; so let us assume that both are nonzero. Then the lowest order term in $\Delta_1$ is $-3H_0q_0^2$ times one of the three terms $h_j^4\tau^{4j}$ or $-2H_0h_j^2z_k\tau^{2j+k}$ or $H_0^2z_k^2\tau^{2k}$, which gives in turn $H_0q_0=0$.\\

\noindent
{\em Intrinsic Singularities of the $I_1$-Surface: The Curve of Higher Double Points}

Besides the cusp curve $C$ the surface $D_1$ has also a curve of higher double points (resolved by a process of three blow-ups)
which turns out to be just the curve $(h)$.
Equ.~(\ref{SU(5) discriminant}) shows that the defining polynomial
for $D_1$ (given in the curly brackets) can be written near $h$ as follows
(keeping for each $z$-power just the leading $h$-power)
\beqa
\label{sing}
-3Hq^2\Big( h^4-2h^2Hz+H^2z^2\Big) + {\cal O}(z^3)
\eeqa
We want to look at the leading terms near $(h,z)=(0,0)$.
Written in the variable $w:=Hz-h^2$ the terms up to third order here
become structurally (i.e.~everything up to coefficients and where $H\neq 0$) $w^2+z^3 \rightarrow h^6+h^4w+w^2$ near $(h,w)=(0,0)$; this gives the normal form $h^6+v^2$ with $v:=w+\frac{3}{2}h^4$, i.e.~the curve $(h)$ is actually a singular curve
of higher double points of $D_1$. So this matter curve does not arise
in the standard framework of the collision rules where two smooth surfaces
intersect transversally.
Note that the prefactor in equ.~(\ref{sing}) shows that
at the points of $(h)$ of $E_6$ and $D_6$ enhancements, where in addition to $h$ also $H$ or $q$ vanishes, respectively, the singularity of $D_1$ will be even worse.\\

\noindent
{\em Intrinsic Singularities of the $(P)$-Curve: Its Double Point Locus}

As equ.~(\ref{P def}) shows the locus $(h)\cap (q)$
of points $p$ of $D_6$-enhancement lies in $(P)$.
If we consider $h$ and $q$ as local functions in $B_2$ near $p$ we find from equ.~(\ref{P def}) the double point structure $q^2+qh+h^2$.
This holds, strictly speaking, only for the case where $(h)$ is {\em not} turned off by going to the pseudo-separation case $t=c_1$,
as then a constant $h$ cannot serve as a local coordinate.\\

\noindent
{\em Comparison with Heterotic String Theory}

For the case that $B_3$ is ${\bf P}^1$ fibered over $B_2$ one can compare with
the heterotic side [\ref{FMW}]. There a vector bundle $V$ of structure group $H$ is specified which breaks the gauge group $E_8$ (we may assume that the second $E_8$ is completely broken). If $H=SU(N)$ for $N=3,4$ or $5$ what remains in the effective fourdimensional gauge theory is the commutator $E_6$, $SO(10)$ or $SU(5)$. The heterotic Calabi-Yau space $Z$ is elliptically fibered
over the surface $B_2$ which is visible to both sides of the duality, i.e.~the
duality arises by expanding adiabatically the eightdimensional duality.
On a generic elliptic fiber the $SU(N)$ bundle decomposes as a sum of line
bundles; each of these is characterized by a fiber point. Globally over $B_2$ these points trace out a surface, the spectral cover $C$. Cohomologically $C=N\sigma+\eta$ where $\sigma$ is the cass of the base $B_2$ and $\eta$ is (the pullback of) a class in $B_2$. The dictionary to the $F$-theory side is implemented by setting $\eta=6c_1-t$. The equation for $C$ is given (in affine fiber coordinates) for an $SU(5)$ bundle by \beqa
a_0+a_2x+a_3y+a_4x^2+a_5xy&=&0
\eeqa
(here the $a_i$ are certain sections over $B_2$ of cohomology class $\eta-ic_1=(6-i)c_1-t$).
The matter localized on curves in $B_2$ arises as follows. In $F$-theory the ${\bf 10}$ arose on the $SO(10)$ enhancement curve $(h)$, and the ${\bf \overline{5}}$ similarly from the $SU(6)$ enhancement curve $(P)$; in the heterotic theory a larger gauge group means having a reduced structure group. This means, for $(h)$, that one of the fiber points of $C$ becomes zero (in the group law), which takes place where the surface $C$ intersects $B_2$:
this happens at the curve defined by $a_5=0$ where the structure
group is reduced to $SU(4)$. So $h$ corresponds to $a_5$ which has the right cohomological degree $c_1-t$.
Similarly the ${\bf 5}$ is supported on the curve $(R)=\{R=0\}$
for the resultant $R=Res(a_0+a_2x+a_4x^2, a_3+a_5x)
= a_0a_5^2-a_2a_3a_5+a_3^2a_4$, corresponding to $P$ in equ.~(\ref{P def}) and of the right cohomological degree $8c_1-3t$.
The complete dictionary is (here the degrees match)
\beqa
\label{h correspondence}
h_{c_1-t} &=&a_5\\
-3 H_{2c_1-t}&=&a_4\\
q_{3c_1-t}&=&a_3\\
3 f_{4c_1-t}&=&a_2\\
2 g_{6c_1-t}-3f_{4c_1}H&=&a_0
\eeqa
Gauge group enhancements to $E_6, SO(12)$ and $SU(7)$ are localized at the intersections $(h)(H)$, $(h)(q)$ and $(P)(Q)$
in $F$-theory, that is, on $(a_5)(a_4)$, $(a_5)(a_3)$ and $(R)(S)$
where $S_{9c_1-3t}=-(\frac{3}{4}a_2^2+2a_0a_4)a_5-a_3^3$. At $\{a_5=a_4=0\}$
the structure group $H$ is reduced to $SU(3)$ with commutator $G=E_6$;
a corresponding reasoning can be applied to the other points.

The results on $e(\overline{X})$ can be compared [\ref{AC}]
via the relation $n_3=n_5$ with a corresponding heterotic computation
where an $SU(N)\x E_8$ bundle $(V_1,V_2)$ is given
\beqa
24\, n_5&=&288+(1200+107N-18N^2+N^3)c_1^2+(1080-36N+3N^2)c_1t
+(360+3N)t^2\;\;\;\;\;\;\\
\label{SU(5) case n5}
&\rightarrow&288+1410c_1^2+975c_1t+375t^2\\
\label{n5 number}
&\rightarrow&288+2760c_1^2
\eeqa
with $N = 0,2,3,4,5$ for a gauge group $G = E_8, E_7, E_6, SO(10), SU(5)$, where we also indicated the specialisations for $SU(5)$ and for the pseudo-separation case $t=c_1$. In the separation case (pseudo-separation for $N=5$) of $t=(6-N)c_1$ we get that the important expression, $\eta -N c_1=(6-N)c_1-t$, related to heterotic matter, cf.~[\ref{A98}, \ref{C}], vanishes.

\section{\label{tadpole section}Tadpole Cancellation}

If $D$-brane charges do not cancel and leave a net RR charge in the vacuum
a tadpole arises. This tadpole is not seen in local models as any excess RR charge can escape to infinity;
however, in global models this issue cannot be ignored.
As the cancellation condition involves an Euler number computation
the considerations about globally consistent packages of special degeneration loci in various codimensions have an application here.

An $F$-theory background contains a number of spacetime filling D3-branes which
are located at points in $B_3$. The condition contains further contributions from supersymmetric fluxes associated to either the bulk supergravity fields on $B$ or to the worldvolume gauge fields on $\Delta$. Tadpole cancellation requires the various contributions to satisfy the condition
\beqa
\frac{e(X)}{24}&=&n_3+\frac{1}{2}\int_X G\wedge G+\int_{D_i} c_2(E_i).
\eeqa
This formula applies for $X$ being smooth.
Which formula has to be applied for the physically relevant case
of a singular fourfold has not been worked out fully.
Also the meaning of the Euler number, if in the correct formula still relevant,
has to be clarified.
We will restrict to the computation of the Euler characteristic of the resolved Calabi-Yau fourfold assuming that such a resolution exists globally and arises from resolving fiberwise (inserting the Hirzebruch trees for the respective Kodaira singularity type of the fiber); clearly this is only the simplest assumption
(this computation agrees in some cases where it can be checked with a toric computation).
The Euler number of a resolved model might still get correction terms in its
function as a contribution to $n_3$; nevertheless we find agreement with a corresponding heterotic computation for the number of fivebranes in some cases which can be compared. In any case this serves as an illustration of the phenomena which occur in considering the contributions of the various relevant subloci.

It is instructive to recall first the {\it smooth case}, i.e.~$X$ has no singularities and is described by a smooth Weierstrass model. As a second step, we consider the case where $X$ only develops an $ADE$ singularity along a codimension-one subvariety $r$ (in itself a problematic assumption as we recalled above in connection with such 'separation cases' and as we will develop further below).
The general case will be discussed again for our $SU(5)$ model.
\vskip 0.5cm
\noindent
{\em Smooth Case}

For a smooth $X$ we expect only contributions from type ${I}_1$ singular fibers over $D_1$ minus $C$ and type ${II}$ singular fibers over the cusp curve $C$
\beqa
e(X)&=&e(I_1)\Big(e(D_1)-e(C)\Big)+e(II)e(C)\;
=\; 288+360c_1(B_3)^3
\eeqa
where $e(D)$ is
the Euler characteristic of the $I_1$ surface, which itself is singular along the cusp curve $C$; we get [\ref{AC}] a ''Pl\"ucker-like'' formula \beqa
e(D)=c_2(B_3)D-c_1(B_3)D^2+D^3+\Delta_C
\eeqa
where $\Delta_C$ is understood as a correction term to the smooth case, i.e.~to the Euler characteristic of a smooth surface inside $B_3$ and we find $\Delta_C=2(e(C)-DC)$ [\ref{AC}] but $\Delta_C$ receives further corrections when specifying a section of $G$ singularities along $r$ (cf.~below). The expression
for the Euler characteristic of $C$ can be easily evaluated in the smooth case by simply noting that
$C=FG$ and using the fact that the normal bundle of $C$ in $B_3$ is given by $N_C\vert_{B_3}=({\cal O}(F)\oplus {\cal O}(G))\vert_C$ and restricting the short exact sequence $0\to T_D\to T_{B_3\vert_D}\to N_{D\vert_{B_3}}\to 0$ to $C$, we find
\beqa
e(C)=c_1(B_3)FG-(F+G)FG
\eeqa
\vskip 0.5cm
\noindent
{\em Singular Case - Codimension-One}

In [\ref{AC}] an Euler number formula was derived for the case with singularities only over the codimension-one locus $B_2$ (with $r_G$ and $c_G$ the rank and Coxeter number, resp., of the gauge group $G$).
Applying the stratification method as above we get
\beqa
\label{shit2}
e(\overline{X})&=&288+360\int_{B_3}c_1^3(B_3)-r_Gc_G(c_G+1)\int_{B_2}
c_1^2(B_2) \\
&\rightarrow&288+\Big(180(12+n^2)-r_Gc_G(c_G+1)\Big)\; c_1^2 \eeqa
($\overline{X}$ the fiberwise resolved model).
Here we indicated also the spezialization
$B_3={\bf F_{k;m,n}}$ over $B_2={\bf F_k}$ with the ${\bf P^1}$ fibration $t=mb+nf$, with actually $k=0$ and $m=n$ from
the pure codimension-one (separation) condition.
The case of having a degeneration purely in codimension-one, i.e.,~a 'separation case' between the two discriminant components $B_2$ and $D_1$ (without matter curve) is realizable for $G=E_8, E_7, E_6, D_4$ over $B_2=F_0$ with $n=m=12,8,6,4$.

Thereby we find agreement between equ.~(\ref{SU(5) case n5}) and [\ref{KLRY}] over $B_2={\bf F_0}$
for true separation cases $E_k$ and $D_4$. For the pseudo-separation cases we find also agreement with our formula. Via a computation using toric geometry and computer analysis [\ref{KLRY}]
one finds the following table \begin{center}
\begin{tabular}{|c|c|}
\hline
$G$   & $e(\overline{X})$   \\
\hline
\hline
$D_4$ & $288+4872c_1^2$\\
\hline
$E_6$ & $288+7704c_1^2$ \\
\hline
$E_7$ & $288+11286c_1^2$\\
\hline
$E_8$ & $288+20640c_1^2$\\
\hline
\end{tabular}
\end{center}
In the $A$ series the pseudo-separation specialization
$t=c_1$ is used such that the $h$ curve is 'turned off' cohomologically and only the $P$ curve remains.

\vskip 0.5cm
\noindent
{\em General Case - $SU(5)$ Singularity}

To illustrate the general case (i.e.~where no special choice of $t$ is made and all matter curves contribute as well as the codimension-three singularities in $B_3$ are present)
we will consider the example of having  an $SU(5)$ singularity (in the fiber)
along $r=B_2$ in $B_3$.
Following the procedure above, we decompose the discriminant $D$ into $D_1+D_2$ where again $D_1$ denotes the component with $I_1$ fibers. With $D_2=5r, F_2=0$ and $G_2=0$ we get from equ.~(\ref{divisor1 equations}) expressions for $D_1, F_1$ and $G_1$.
To determine the Euler characteristic of $\overline{X}$ we have to compute first the Euler characteristic of the singular surface $D_1$ taking its singularity structure into account. We showed that $D_1$ is singular along the cusp curve $C$ and also along a curve of higher double points which we identified as the matter curve $h$. Thus when computing the Euler characteristic of $D_1$ we actually expect correction terms $\Delta_C$ and $\Delta_h$ to $e(D_1^{\rm smooth})$. Now as the cusp and higher double point curve intersect we expect a further correction term $\Delta_{C\cap h}$. Moreover, the singularity structure along $h$ will change if at special loci coefficient functions vanish so that we get degenerations of the structure of equ.~(\ref{sing}) which happens at the loci $h\cap q$ and $h\cap H$ (we also include a term $\Delta_p$ corresponding to possible corrections from other point singularities of $D_1$). In summary, we get
\beqa
e(D_1)=c_2(B_3)D_1-c_1(B_3)D_1^2+D_1^3+\Delta_C+\Delta_h+\Delta_{C\cap h}
+\Delta_{h\cap q}+\Delta_{h\cap H}+\Delta_p
\eeqa
where $\Delta_C=2(e(C)-CD)+25Cr$ (here one has to work with $C$ redefined by equ.~(\ref{Cusp});
for the expressions of the other correction terms $\Delta_i$ cf.~[\ref{AC}]). Summarizing all contributions we find
(where $\#$ denotes the cardinality of a set and polynomials stand for their zero-divisors) \beqa
e(\overline{X})&=&
+e(I_1)\Big[e(D_1)-e(C)-e(h)-e(P)+\#(h\cap P)+\#(C\cap r)\Big]\nonumber\\
& &+e(II)\Big[e(C)-\#(C\cap r)\Big]\nonumber\\
& &+e(A_4)\Big[e(B_2)-e(h)-e(P)+\#(h\cap P)\Big]\nonumber\\
& &+e(D_5)\Big[e(h)-\#(h\cap P)\Big]+e(A_5)\Big[e(P)-\#(h\cap P)-\Big(\#(P\cap Q)-\#(h\cap q)\Big)\Big]
\nonumber\\
& &+e(E_6)\Big[\#(h\cap H)\Big]+e(D_6)\Big[\#(h\cap q)\Big]
+e(A_6)\Big[\#(P\cap Q)-\#(h\cap q)\Big]\\
&=&e(D_1)+e(C)- \#(C\cap r)
+5 e(B_2)+\#(h\cap P)+\#(P\cap Q)-\#(h\cap q)
\eeqa
(note that $h\cap P = (h\cap H)\cup (h\cap q)$ which is a disjoint decomposition, so the cardinalities add).
If actually $C\cap r=(h)\cap (P)$ one gets
$e(D_1)+e(C)+5 e(B_2)+\Big(\#(P\cap Q)-\#(h\cap q)\Big)$.
This formula can be compared with equ.~(\ref{SU(5) case n5}).
In the general case the singular structure of $D_1$ along $(h)$ has to be taken
into account.
Here we restrict us to the pseudo-seperation case of $t=c_1$ where the matter curve $(h)$ is turned off and $D_1\cap r=(P), C\cap r=\emptyset$. Then \beqa
e(\overline{X})&=&e(D_1)+e(C)+5e(B_2)+\#(P\cap Q)
\eeqa
So for this pseudo-seperation case we find \beqa
e(D_1)&=&c_2(B_3)D_1-c_1(B_3)D_1^2+D_1^3+\Delta_C=228+4493c_1^2
\eeqa
and using $e(B_2)=12-c_1^2$ and $e(C)=\Big(c_1(B_3)-(F_1+G_1)\Big)F_1G_1
=-1728c_1^2$ we find
\beqa
e(\overline{X})=288+2760c_1^2+\#(P\cap Q)
\label{shit}
\eeqa
The $n_3=e(\overline{X})/24$ matches not immediately the corresponding number of heterotic five-branes, cf.~equ.~(\ref{n5 number}). So either there is a further singular contribution (in $e(D_1)$ or $e(C)$)
or here the contribution to the number of three-branes $n_3$ is not derived from the Euler number of the fiberwise resolved model $\overline{X}$. This shows again the special status of the codimension three locus $P\cap Q$ which does not arrive as intersection of matter curves (which themselves arise as intersections of $D7$-brane components). \vskip 0.2cm
\noindent
{\it Hodge Numbers and $e(\overline{X})$}

A comparison of moduli spaces for $F$-theory and a dual heterotic string theory on a Calabi-Yau threefold $Z$, elliptically fibered over $B_2$, gives the following expressions [\ref{AC97}]
\beqa
h^{11}(\overline{X})&=&h^{11}(Z)+1+rk=12-c_1^2+rk\\
h^{31}(\overline{X})&=&h^{21}(Z)+I+n_o+1=12+29c_1^2+I+n_o\\
h^{21}(\overline{X})&=&n_o
\eeqa
for the Hodge numbers of a resolved $\overline{X}$.
Here $I=I_{SU(N)}+I_{E_8}$ denote the number of moduli of the $SU(N)$ resp.~$E_8$ bundle in the heterotic model with
\beqa
I_{SU(N)}&=&(N-1)+\Big(\frac{N^3-N}{6}-3N^2+18N+6\Big)c_1^2
+\Big(\frac{N^2}{2}-6N-1\Big)c_1t+\frac{N}{2}t^2\\
I_{E_8}&=&8+166c_1^2+181c_1t+60t^2.
\eeqa
Further $n_o$ refers to the total number of odd bundle moduli, $rk=16-(N-1)-8$, and we have $\eta_{SU(N)}=6c_1-t$ and $\eta_{E_8}=6c_1+t$.
The Euler characteristic of $\overline{X}$ can be expressed as [\ref{SVW}]
\beqa
e(\overline{X})=
48+6\Big(h^{11}(\overline{X})-h^{21}(\overline{X})+h^{31}(\overline{X})\Big)
\eeqa
and inserting the above expressions gives
\beqa
e(\overline{X})=
288+\Big(1200+107N-18N^2+N^3\Big)c_1^2+\Big(1080-36N+3N^2\Big)c_1t
+\Big(360+3N\Big)t^2.
\label{n5}
\eeqa
Thus for $N=5$, $t=c_1$ we get $e(\overline{X})=288+2760c_1^2$ matching the heterotic result equ.~(\ref{n5 number}).

\section{The other $SU(n)$ Cases}

Here we give corresponding informations for the other $SU(n)$ cases, cf. appendix and also [\ref{AC}].\\

\noindent
{\em The Locus $C\cap B_2$}

\vspace{.2cm}
In all $I_n$ cases we have $C\cap r\subset (h)$, resp. $\subset (H)$
for $n=2$. For the cohomological intersection one finds the following.
For $n=4,5,6$ where $C=(f)(g)-3n(h)r$ one has again
\beqa
Cr=3[h](8[h]+nt)=3[h][P].
\eeqa Here the (cohomological) degree (i.e., cohomology class [P])
of the second matter curve $P$
is generally computed as follows: from $D=12(c_1+2t+r)$ one has $Dr=12(c_1-t)=12[h]$
and from $D=nr+D_1$ one finds that $Dr=-nt+4[h]+[P]$, or $[P]=8[h]+nt$.

Similarly for $n=2$ and $3$ one gets the following: for $n=2$ where
$(f)|_r=2(H)$, $(g)|_r=3(H)$ and $C=(f)(g)-3(H)r$ one has \beqa
Cr=3[H](2[H]+t)=\frac{3}{2}[H][P],
\eeqa
as one has from $Dr=6[H]$ and $D_1r=2[H]+[P]$
that $6[H]=-2t+2[H]+[P]$, or that $[P]=4[H]+2t$; for $n=3$ where $C=(f)(g)-8(h)r$ one has \beqa
Cr=8[h](3[h]+t)=\frac{8}{3}[h][P]
\eeqa
as one gets from $D_1r=3[h]+[P]$ that $12[h]=-3t+3[h]+[P]$, or $[P]=9[h]+3t$.\\

\noindent
{\em Euler Characteristic for the Pseudo-Separation Case $t=c_1$}

\vspace{.2cm}
The pseudo-seperation case $t=c_1$ gives for the $SU(n)$ series \beqa
e({\overline X})&=&e(I_1)\Big[e(D_1)-e(C)-e(P)\Big]+e(II)e(C)\nonumber\\
& &+e(A_{n-1})\Big[e(B_2)-e(P)\Big]
+e(A_n)\Big[e(P)-\#(P\cap Q)\Big]+e(A_{n+1})\Big[\#(P\cap Q)\Big]\nonumber\\
&=&e(D_1)+e(C)+ne(B_2)+\#(P\cap Q)\nonumber\\
&=&288+\Big(2880-n(n^2-1)\Big)c_1^2+\#(P\cap Q).
\eeqa
This shows a correction $2872c_1^2+\#(P\cap Q)$ relative to the proper separation case of equ.~(\ref{shit2}).
For the discussion of the special role of the locus $P\cap Q$ cf.~the remark after equ.~(\ref{shit}). \begin{center}
\begin{tabular}{|c|c|}
\hline
$G$   & $e({\overline X})-\#(P\cap Q)$ \\
\hline
\hline
$A_1$ & $288+2874c_1^2$ \\
\hline
$A_2$ & $288+2856c_1^2$ \\
\hline
$A_3$ & $288+2820c_1^2$ \\
\hline
$A_4$ & $288+2760c_1^2$ \\
\hline
$A_5$ & $288+2670c_1^2$ \\
\hline
\end{tabular}
\end{center}
Here $e({\overline X})-\#(P\cap Q)$ matches the number of heterotic five-branes,
cf.~[\ref{AC}].

\appendix

\section{Appendix}

Here we collect some useful results for the other $SU(n)$ cases
which parallel those in the main text for our main example $SU(5)$.

\subsection{Case $SU(2)$}
The discriminant has the following structure
\beqa
\label{disc I2}
\Delta=z^2\Big(H^2P+(-f_3P+Q)z+g_4^2z^2\Big)
\eeqa
with \beqa P&=&-\frac{3}{4}f_3^2+2g_4H+3f_2H^2\\
Q&=&f_3\Big(\frac{1}{4}f_3^2-g_4H-3f_2H^2\Big)+2g_3H^3
\eeqa
and \beqa f=\frac{1}{48}(-H^2+f_3z+f_2z^2), \ \ \ g=\frac{1}{864}(H^3-\frac{3}{2}f_3Hz+g_4z^2+g_3z^3)
\eeqa
Let us now read off the various relevant subloci from the
discriminant equation (\ref{disc I2})
\begin{itemize}
\item
{\em Codimension one:} The divisor of the $SU(2)$ singularity, i.e.~the surface $B_2$ of divisor $r$ with multiplicity $2$, is represented
by the factor $z^2$; the other factor is the equation for $D_{I_1}$
\item
{\em Codimension two:} There are two matter curves (singularity enhancements in codimension two in $B_3$)
\beqa H=0 &\Longrightarrow& (a,b,c)=(1,2,3) \;\;\buildrel \wedge \over = \;\; A_1\to A_1 (III)\\
P=0 &\Longrightarrow& (a,b,c)=(0,0,3) \;\;\buildrel \wedge \over = \;\; A_1\rightarrow A_2
\eeqa
\item
{\em Codimension three:} The singularity type is enhanced even further in
codimension three
(the intersection of the matter curves $(H)$ and $(P)$
is here the locus where $H=f_3=0$)
\beqa
H=f_3=0 &\Longrightarrow& (a,b,c)=(2,2,4) \;\;\buildrel \wedge \over = \;\; A_1\rightarrow A_2 (IV)\\
P=Q=0  \;\;\;(\mbox{but not} \; H=f_3=0) &\Longrightarrow& (a,b,c)=(0,0,4) \;\;\buildrel \wedge \over = \;\; A_1\rightarrow A_3
\eeqa
\end{itemize}

\subsection{Case {$SU(3)$}}
Here the discriminant is
\beqa
\label{disc I3}
\Delta=z^3\Big(h^3P+(-qP+Q)z+\Big[h(-3f_2g_3h+6f_2^2q)+3g_3q^2\Big]z^2
+(f_2^3+g_3^2)z^3\Big)
\eeqa
with \beqa P&=&-q^3-3f_2h^2q+2g_3h^3\\
Q&=&\frac{5}{4}q^4+\frac{9}{2}f_2h^2q^2-4g_3h^3q-\frac{3}{4}f_2^2h^4
\eeqa
and
\beqa f=\frac{1}{48}(-h^4+2hqz+f_2z^2), \ \ \ g=\frac{1}{864}\Big(h^6-3h^3q
+\frac{3}{2}(q^2-f_2h^2)z^2+g_3z^3\Big)
\eeqa
Again we read off from (\ref{disc I3}) the various subloci
\begin{itemize}
\item {\em  Codimension one:} The divisor of the $SU(2)$ singularity, i.e.~the surface $B_2$ of divisor $r$ with multiplicity $2$, is represented
by the factor $z^2$; the other factor is the equation for $D_{I_1}$
\item
{\em Codimension two:} Here we have again two matter curves \beqa h=0 &\Longrightarrow& (a,b,c)=(2,2,4) \;\;\buildrel \wedge \over = \;\; A_2\rightarrow A_2 (IV)\\
P=0 &\Longrightarrow& (a,b,c)=(0,0,4) \;\;\buildrel \wedge \over = \;\; A_2\rightarrow A_3
\eeqa
\item
 {\em Codimension three:} The singularity type is enhanced even further in
codimension three (the intersection of the matter curves $(h)$ and $(P)$
is here the locus where $h=q=0$)
\beqa
h=q=0 &\Longrightarrow& (a,b,c)=(2,3,6) \;\;\buildrel \wedge \over = \;\; A_2\rightarrow D_4\\
P=Q=0  \;\;\;(\mbox{but not} \; h=q=0) &\Longrightarrow& (a,b,c)=(0,0,5) \;\;\buildrel \wedge \over = \;\; A_2\rightarrow A_4
\eeqa
(note that the $D_4$ point is a triple or quadruple point
of $P$ or $Q$, respectively).
\end{itemize}

\subsection{Case {$SU(4)$}}
Here the discriminant equation looks as follows
\beqa
\Delta=z^4\Big(h^4P+h^2[-2HP+Q]z+{\cal O}(z^2)\Big)
\eeqa
with (where $e:=f_2+H^2, k:=2g_2-3f_1H$)
\beqa
P&=&h^2k-\frac{3}{4}e^2\\
Q&=&-h^2\Big(kH-2g_1h^2+\frac{3}{2}f_1e\Big)
\eeqa
and
\beqa
f&=&\frac{1}{48}\Big(-h^4+2h^2Hz+f_2z^2+f_1z^3\Big)\\
g&=&\frac{1}{864}\Big(h^6-3h^4Hz+\frac{3}{2}h^2(H^2-f_2)z^2
+\Big[\frac{1}{2}H(H^2+3f_2)-\frac{3}{2}f_1h^2\Big]z^3+g_2z^4+g_1z^5\Big)\;\;\;
\eeqa
Now we read off the various subloci
\begin{itemize}
\item {\em  Codimension one:} The divisor of the $SU(4)$ singularity, represented by the factor $z^4$, is the surface $B_2$ of divisor $r$ with multiplicity $2$; the other factor is the equation for $D_{I_1}$
\item
{\em Codimension two:} We have again two matter curves \beqa h=0 &\Longrightarrow& (a,b,c)=(2,3,6) \;\;\buildrel \wedge \over = \;\; A_3\rightarrow D_4\\
P=0 &\Longrightarrow& (a,b,c)=(0,0,5) \;\;\buildrel \wedge \over = \;\; A_3\rightarrow A_4
\eeqa
\item {\em Codimension three:} Finally the further singularity enhancement
in codimension three (the intersection of the matter curves $(h)$ and $(P)$
is here the locus where $h=f_2=0$)
\beqa
h=f_2=0  &\Longrightarrow& (a,b,c)=(3,3,6) \;\;\buildrel \wedge \over = \;\; A_3\rightarrow D_4\\
h=H=0 &\Longrightarrow& (a,b,c)=(2,4,6) \;\;\buildrel \wedge \over = \;\; A_3\rightarrow D_4\\
P=Q=0 &\Longrightarrow& (a,b,c)=(0,0,6) \;\;\buildrel \wedge \over = \;\; A_3\rightarrow A_5
\eeqa
\end{itemize}

\section{References}

\begin{enumerate}

\item
\label{V1}
  C.~Beasley, J.~J.~Heckman and C.~Vafa,
 {\em GUTs and Exceptional Branes in F-theory - I,}
  JHEP {\bf 0901} (2009) 058
  arXiv:0802.3391 [hep-th].\\
  C.~Beasley, J.~J.~Heckman and C.~Vafa,
  {\em GUTs and Exceptional Branes in F-theory - II: Experimental Predictions,}
  JHEP {\bf 0901} (2009) 059
  arXiv:0806.0102 [hep-th].

\item
\label{BIKMSV}
M. Bershadsky, K. Intriligator, S. Kachru, D.R. Morrison, V. Sadov and C. Vafa,
{\em Geometric Singularities and Enhanced Gauge Symmetries},
hep-th/9605200, Nucl.Phys. {\bf B481} (1996) 215.

\item
\label{SVW}
S. Sethi, C. Vafa and E. Witten, {\em Constraints on Low Dimensional String Compactifications}, hep-th/9606122, Nucl. Phys. {\bf B480} (1996) 213.

\item
\label{KLRY}
A. Klemm, B. Lian, S.S. Roan and S.T. Yau,
{\em Calabi-Yau fourfolds for M- and F-Theory compactifications},
hep-th/9701023, Nucl.Phys. {\bf B518} (1998) 515.

\item
\label{AC97}
B. Andreas and G. Curio,
{\em Three-branes and five-branes in N=1 dual string pairs,} hep-th/9706093, Phys.Lett. {\bf B417} (1998) 41-44.

\item
\label{AC}
B. Andreas and G. Curio,
{\em On discrete Twist and Four-Flux in N=1 heterotic/F-theory compactifications},
hep-th/9908193, Adv.Theor.Math.Phys. {\bf 3} (1999) 1325.

\item
\label{FMW}
R.~Friedman, J.~Morgan and E.~Witten, {\em Vector Bundles And F Theory},
 arXiv:hep-th/9701162, Commun.Math.Phys. 187 (1997) 679

\item
\label{A98}
B.~Andreas,
{\em On vector bundles and chiral matter in N = 1 heterotic
compactifications,} hep-th/9802202, JHEP {\bf 9901} (1999) 01.

\item
\label{C}
G. Curio,
{\em Chiral matter and transitions in heterotic string models},
hep-th/9803224, Phys.Lett. B435 (1998) 39.

\item
\label{Aluffi}
P. Aluffi and M. Esole,
{\em Chern class identities from tadpole matching in type IIB and F-theory},
arXiv:0710.2544.

\item
\label{DW}
R.~Donagi and M.~Wijnholt,
{\em Model Building with F-Theory},
 arXiv:0802.2969.\\
R.~Donagi and M.~Wijnholt,
{\em Breaking GUT Groups in F-Theory}, arXiv:0808.2223.

\item
\label{HKTW}
H.~Hayashi, T.~Kawano, R.~Tatar and T.~Watari,
{\em Codimension-3 Singularities and Yukawa Couplings in F-theory},
 arXiv:0901.4941.

\end{enumerate}

\end{document}